\documentstyle{depart}
\begin{document}
\baselineskip22pt

\begin{center} Modification of the Doppler Effect due to \\ the Helicity -- Rotation Coupling\\

\vspace{.25in}

Bahram Mashhoon\\ Department of Physics and Astronomy\\ University of Missouri - Columbia\\ Columbia,
Missouri 65211, USA\\

\vspace{.75in}

Abstract\\
\end{center}
\vspace{.03in}

The helicity-rotation coupling and its current empirical basis are examined.  The modification of the
Doppler effect due to the coupling of photon spin with the rotation of the observer is considered in
detail in connection with its applications in the Doppler tracking of spacecraft.  Further implications
of this coupling and the possibility of searching for it in the intensity response of a rotating
detector are briefly discussed. 

\vspace{.15in}


\newpage

\noindent 1.  Introduction
\vspace{.05in}

The standard relativistic Doppler and aberration formulas can be derived from the invariance of the
phase of the radiation under Lorentz transformations [1].  This phase is not in general an invariant,
however, when the transformations connect noninertial frames of reference.  In this case, the Doppler
and aberration formulas need to be modified by terms that depend on the accelerations involved [2, 3]. 
These modifications are usually very small and disappear in the high-frequency limit; that is, the
standard relativistic formulas are recovered in the ray limit $\lambda/{\cal L}\rightarrow 0$, where
$\lambda$ is the wavelength of the radiation and ${\cal L}$ is the effective acceleration length
associated with the observer or the measuring device that is used for observation.  For instance, ${\cal
L}=c^2/a$ or $c/\Omega$ for the translational acceleration
$a$ or the rotational frequency $\Omega$ of the observer or the measuring device; to avoid this
repetition, the term ``observer'' is henceforth employed in an extended sense to include any
appropriate measuring device.  The most significant effect of this type is due to the coupling between
the helicity of the radiation and the rotation of the observer, since it reveals the inertia of intrinsic
spin [4, 5].

Consider a global inertial frame with an ideal set of fundamental static inertial observers.  According
to these observers, a plane monochromatic electromagnetic wave has frequency $\omega$ and wave vector
${\bf k}$.  We are interested in the reception of this wave by a noninertial observer that moves with
velocity ${\bf v}(t)$ and refers its observations to a local orthonormal spatial triad along its path
that rotates with frequency $\mbox{\boldmath$\Omega$}(t)$ as measured by the fundamental observers with
respect to their spatial axes.  According to the noninertial observer, the frequency and wave vector of
the electromagnetic wave are given in the eikonal approximation, i.e. for $\omega>>\Omega$, by 

\begin{eqnarray}
\omega^{\prime}&=&\gamma[(\omega-{\bf\hat H}\cdot\mbox{\boldmath$\Omega$})-{\bf v}\cdot{\bf k}]\;\;,\\
{\bf k}^{\prime} &=& {\bf k} +\frac{1}{v^2} (\gamma - 1)({\bf v}\cdot{\bf k}){\bf v} -
\frac{1}{c^2}\gamma(\omega - {\bf\hat H}\cdot\mbox{\boldmath$\Omega$}){\bf v}\;\;,
\end{eqnarray}

\noindent where ${\bf\hat H} =\pm c{\bf k}/\omega$ is the unit helicity vector of the wave.  The upper
(lower) sign in ${\bf\hat H}$ indicates that for the fundamental observers facing the arriving wave the
electric and magnetic fields rotate with frequency $\omega$ in the counterclockwise (clockwise) sense
along the direction of propagation of the wave.  That is, ${\bf\hat H} = {\bf\hat k}$ implies that the
beam is right circularly polarized (RCP) and has positive helicity, while ${\bf\hat H} = -{\bf\hat k}$
implies that the beam is left circularly polarized (LCP) and has negative helicity.  We assume that
during the time (typically a few periods of the wave) that it takes to determine the frequency
$\omega^{\prime}$ and wave vector ${\bf k}^{\prime}, {\bf v}$ and $\mbox{\boldmath$\Omega$}$ are
effectively constants, so that in this sense their variation over time can be considered adiabatic. 
Equations (1) and (2) represent the modified expressions for the Doppler effect and aberration for a
rotating observer [3].  Various consequences of the modified equations for the aberration of polarized
radiation and for interferometry with polarized radiation in a rotating frame of reference have been
discussed in [3].  Therefore, the present Letter is devoted to the modified Doppler effect, as new
observations are sensitive to the polarization dependence of the modified Doppler formula (1).  
\vspace{.25in}

\noindent 2. Modified Doppler effect
\vspace{.05in}

The Doppler effect is important in many branches of basic physics such as spectroscopy, astrophysics and
cosmology; moreover, it has major practical applications in remote sensing and navigation.  In
particular, in connection with communication with artificial satellites (e.g., DSN, GPS), we note that
circularly polarized radio beams are generally employed [6-9].  Ignoring the correction due to the
helicity-rotation coupling in (1) and employing the standard (relativistic) Doppler formula would
introduce an error in the velocity determination given by $|\Delta{\bf v}|\sim c\Omega/\omega$.  On the
other hand, ignoring the helicity-rotation coupling in equation (2) would result in a {\it relative}
error in the aberration angle of $\sim\Omega/\omega$.  Let us note for the sake of reference that for
the GPS radio waves with frequency $\sim 1{\rm GHz}, \Omega_{\oplus}/\omega\sim 10^{-14}$ for the
rotation of the Earth about its axis, while for a receiver rotating at $\sim 10$ cps we have
$\Omega/\omega\sim 10^{-8}$. 

To elucidate equation (1), we note that it consists of the time dilation effect represented by the
Lorentz $\gamma$-factor, the linear (Galilean) Doppler effect and the helicity-rotation coupling
effect.  We concentrate on the last effect, which is clearly not as well known as the standard Doppler
effect.  To this end, imagine a rotating observer with its center of mass at rest at the origin of
spatial coordinates in the global inertial reference frame.  The observer refers its observations to
axes that rotate with frequency
$\Omega$ about the $z$-axis, i.e.

\begin{eqnarray}   {\bf\hat x}^{\prime} &=&{\bf\hat x}\;\cos\Omega t + {\bf\hat y}\;\sin\Omega t\;\;,\\
{\bf\hat y}^{\prime} &=& -{\bf\hat x}\;\sin\Omega t + {\bf\hat y}\;\cos\Omega t\;\;,
\end{eqnarray}

\noindent and ${\bf\hat z}^{\prime} = {\bf\hat z}$.  A plane monochromatic circularly polarized
electromagnetic wave propagates in the inertial frame along a direction ${\bf\hat k}$, which may be
chosen to be ${\bf\hat k} = - \sin\theta\;{\bf\hat y} + \cos\theta\;{\bf\hat z}$, where $\theta$ is a
polar angle, with no loss in generality.  The electric field of the wave may be expressed as 

\begin{equation}  {\bf E}(t, {\bf r}) = {\rm Re}[A({\bf\hat x}\pm i {\bf\hat n}) \;\;e^{-i\omega(t -
{\bf\hat k}\cdot{\bf r})}]\;\;,
\end{equation}

\noindent where $A$ is a constant complex amplitude and ${\bf\hat n} = {\bf\hat k}\times{\bf\hat x} =
\cos\theta\;{\bf\hat y} + \sin \theta\;{\bf\hat z}$ is such that $({\bf\hat x}, {\bf\hat n}, {\bf\hat
k})$ form an orthonormal triad.  The corresponding magnetic field is given by ${\bf B} = {\bf\hat
k}\times {\bf E}$; therefore, in what follows we only need to deal with the electric part of the
radiation field.  The upper (lower) sign in equation (5) refers to positive (negative) helicity
radiation.  The components of the electric field as measured by the noninertial observer at
${\bf r} = 0$ are

\begin{eqnarray}  {\bf E} \cdot {\bf\hat x}^{\prime} &=& {\rm Re} [A(\cos\Omega t\pm
i\cos\theta\sin\Omega t)e^{-i\omega t}]\;\;,\\  {\bf E} \cdot {\bf\hat y}^{\prime} &=& {\rm Re} [\pm
iA(\cos\theta\cos\Omega t \pm i\sin\Omega t)e^{-i\omega t}]\;\;,\\  {\bf E}\cdot {\bf\hat z}^{\prime}
&=& {\rm Re} (\pm i A\sin\theta\;e^{-i\omega t})\;\;.
\end{eqnarray}

The Fourier analysis of equations (6) - (8) for $t: -\infty\rightarrow\infty$ reveals their frequency
content $\omega^{\prime} = \omega - m\Omega$, where $m=0,\pm 1$.  The integer $m$ has a simple physical
interpretation here : The component of photon spin along the direction of rotation of the
observer---taken to be the axis of quantization---is $\hbar m$; that is, $\hbar$, 0 or $-\hbar$.  The
resulting three frequencies $\omega - \Omega, \omega$ and $\omega + \Omega$ occur with different
amplitudes that can be simply calculated from equations (6) - (8); for this purpose, we consider the
complex field ${\bf E}^c$, where ${\bf E} = {\rm Re}({\bf E}^c)$.  For an incident wave of definite
helicity, the complex amplitudes $\psi^{\prime}$ of $\omega - \Omega$ are ${1\over 2}A (1\pm\cos\theta)$
in equation (6) and ${1\over 2} iA(1\pm\cos\theta)$ in equation (7).  Moreover, the complex amplitude of
$\omega$ is
$\pm iA\sin\theta$ in equation (8).  Finally, the complex amplitudes of $\omega + \Omega$ are ${1\over
2}A(1\mp\cos\theta)$ in equation (6) and $-{1\over 2}iA(1\mp\cos\theta)$ in equation (7).  These results
cannot be directly connected with the Doppler effect, since the standard Doppler formula assigns a single
frequency to the rotating observer for a given frequency of incident radiation.  The situation here is
analogous to the semi-classical approximation in wave mechanics.  To establish a connection with the
Doppler formula in the eikonal approximation, we need to define an appropriate {\it average} frequency
measured by the rotating observer.  Following the wave-mechanical analogy, we let the average frequency
be $<\omega^{\prime}> = (\Sigma\omega^{\prime}|\psi^{\prime}|^2)/(\Sigma|\psi^{\prime}|^2)$, where the
summations (containing five terms each) are over the three spatial components of the complex field and
involve the five amplitudes mentioned above.  A simple calculation shows that $\Sigma|\psi^{\prime}|^2
= 2|A|^2$ and $\Sigma\omega^{\prime}|\psi^{\prime}|^2 = 2|A|^2(\omega\mp\Omega\cos\theta)$, so that
$<\omega^{\prime}> = \omega - {\bf\hat H}\cdot\mbox{\boldmath$\Omega$}$.  This result will be proved
directly using the eikonal approximation in section 3.

Let us note that for $\theta=0, {\bf E}\cdot{\bf\hat x}^{\prime} = {\rm Re}[A\;{\rm
exp}(-i\omega^{\prime}t)], {\bf E}\cdot {\bf\hat y}^{\prime} = {\rm Re}[\pm i A\;{\rm
exp}(-i\omega^{\prime} t)$ and ${\bf E}\cdot {\bf\hat z}^{\prime} = 0$, where $\omega^{\prime} =
\omega\mp\Omega$.  Intuitively, the noninertial observer perceives radiation of the same helicity but
different frequency; that is, when the observer rotates in the same (opposite) sense as the electric and
magnetic fields, these appear to rotate with frequency $\omega - \Omega\:(\omega +\Omega)$.  The
frequency measured by the noninertial observer is then $\omega^{\prime} = \omega - {\bf\hat
H}\cdot\mbox{\boldmath$\Omega$}$.  We may express this result in terms of the photon energy as ${\cal
E}^{\prime} = {\cal E} - {\bf S}\cdot\mbox{\boldmath$\Omega$}$, where ${\bf S} = \hbar{\bf\hat H}$ is
the photon spin.  This interpretation in terms of helicity-rotation coupling can be checked for
$\theta=\pi$, since it follows from equations (6) - (8) that $\omega^{\prime} = \omega\pm\Omega$.  These
results, for the case that the axis of rotation corresponds to the direction of wave propagation, are
supported by observational data in the radio, microwave and optical domains.  It is interesting to
present this evidence in some detail.

In the case of radio waves, direct observational evidence for the helicity-rotation coupling exists in
terms of the GPS phase wrap-up [8].  According to [8], the helicity-rotation coupling effect (``phase
wrap-up'') has been measured for GPS signal of frequency $\sim 1 {\rm GHz}$ with a receiver antenna
rotating at a frequency of 8 cps.  For microwaves, a certain rotational frequency shift that depends on
the helicity of the microwave radiation was first discovered by Allen [10], who extended the work of Beth
[11, 12].  The angular momentum carried by circularly polarized light was first directly measured by
Beth [11, 12] and subsequently by a number of investigators [10].  Allen performed experiments using a
thin half-wave conducting dipole that was suspended transversely in a dominant-mode circular waveguide. 
The torque exerted by the circularly polarized microwaves on the dipole could set it into continuous
rotation reaching a terminal uniform rate of 20 cps.  Allen [10] then discovered that circularly
polarized radiation passing through such a rotating helicity flipper emerges with its frequency shifted
by twice the rotation frequency of the helicity flipper.  This general phenomenon is a consequence of the
helicity-rotation coupling [13] and can be simply explained as follows:  For RCP radiation of frequency
$\omega_{\rm in}$ incident normally on a uniformly rotating helicity flipper (e.g., a half-wave plate in
the optical case), the frequency as measured in the flipper is constant and is given by $\omega^{\prime}
=\omega_{\rm in} - \Omega$.  The emerging radiation of frequency $\omega_{\rm out}$ is LCP and its
relation with $\omega^{\prime}$ is $\omega^{\prime} = \omega_{\rm out} + \Omega$.  Therefore,
$\omega_{\rm out} - \omega_{\rm in} = -2\Omega$, so that the down-shift in frequency is twice the
rotation frequency of the flipper.  A similar analysis leads to an up-shift in the case of
LCP$\rightarrow$ RCP.  Allen [10] pointed out that these frequency shifts are analogous to the spectral
shifts associated with the rotational Raman effect in the case of linear molecules.  This analogy was
further studied by Newburgh and Borgiotti [14] who investigated the reflection of microwave radiation
from rotating short wires. 

In the optical domain, the up/down frequency shifter based on the helicity-rotation coupling has been
used in heterodyne interferometry by Crane [15] and further explored by a number of investigators
[16-20].  Garetz and Arnold [21] demonstrated the frequency shift experimentally and discussed
rotation-induced optical activity.  Later, Garetz [22] discussed further applications such as the
measurement of rotational motion of small particles.  The connection between the helicity-rotation
frequency shift and the Pancharatnam and Berry phases has been explored by Simon, Kimble and Sudarshan
[23], Bretenaker and Le Floch [24] and Bhandari [25].  The change in the energy of photons passing
through rotating anisotropic elements has been investigated in detail via simple experiments by Bagini
et al. [26] and further elucidated theoretically by Pippard [27].
\vspace{.25in}

\noindent 3. Oblique incidence
\vspace{.05in}

Let us now return to equations (6) - (8) and consider the case of oblique incidence.  It is instructive
to treat the case of normal incidence first.  For $\theta =\pi/2$, we find ${\bf E}\cdot{\bf\hat
x}^{\prime} = {\rm Re}[A\cos\Omega t\;{\rm exp}(-i\omega t)], {\bf E}\cdot{\bf\hat y}^{\prime} =
{\rm Re}[-A\sin\Omega t\;{\rm exp}(-i\omega t)]$ and ${\bf E}\cdot{\bf\hat z}^{\prime} = {\rm Re}[\pm i
A\;{\rm exp}(-i\omega t)]$, so that only the last term of frequency $\omega$ changes sign when we change
from RCP to LCP radiation, while the first two terms contain frequencies $\omega -\Omega$ and $\omega
+\Omega$ in the rotating frame.  It follows from the wave-mechanical analogy in section 2 that on the
average $<\omega^{\prime}> = \omega$ in this case, a result that is expected to hold in the eikonal
approximation.  To develop such an approximation scheme, it turns out that we must discuss the time it
would take for the noninertial observer to perform measurements.  This circumstance follows from the
basic assumption that is employed in the special theory of relativity in order to extend the physics of
Lorentz invariance to all accelerated observers :  A {\it noninertial} observer is at each instant of
time equivalent to an otherwise identical momentarily comoving {\it inertial} observer.  This {\it
hypothesis of locality} is exact if physical phenomena could all be expressed in terms of pointlike
coincidences of classical particles and electromagnetic rays of radiation.  Indeed, the hypothesis of
locality originates from Newtonian mechanics, where the noninertial observer and the instantaneously
comoving inertial observer share the same state (i.e., position and velocity) and are therefore
equivalent.  Hence, the description of accelerated frames of reference in Newtonian mechanics does not
require the introduction of any new physical hypothesis.  However, such a hypothesis is needed in the
presence of classical wave phenomena that have intrinsic scales of length $(\lambda)$ and time
$(\omega^{-1})$.  The point is that a noninertial observer is endowed with an invariant timescale given
by the acceleration time ${\cal L}/c$.  To measure wave properties, at least a few periods of the wave
must be observed.  If the time required for such an elementary measurement is negligibly small compared
to ${\cal L}/c$, then the noninertial observer is in effect momentarily inertial.  Connecting this local
inertial frame of the observer with the global background inertial frame, we can then derive the
standard Doppler and aberration formulas based on phase invariance.  This would correspond in the case
under consideration to the eikonal {\it limit} $\Omega/\omega\rightarrow 0$ and $\Omega t\rightarrow 0$,
where we have assumed, without any loss in generality, that the measurement of wave properties by the
rotating observer begins at $t=0$.  However, to derive the helicity-rotation coupling in equation
(1), we need to assume that $\omega^{-1}<< t <<\Omega^{-1}$.  This intermediate regime in the
measurement process corresponds to the intermediate character of equation (1); that is, for almost
instantaneous observation, the result can only be consistent with the ray limit $\lambda/{\cal
L}\rightarrow 0$ of equation (1), which is the standard relativistic Doppler formula, while for
observation over a very long time interval $(t>>\Omega^{-1}$ and $t>>\omega^{-1})$ the result follows
from the Fourier analysis of the measured field, $\omega^{\prime}=\gamma(\omega - M\Omega)$, where $M=0,
\pm 1, \pm 2$,... involves the total (orbital plus spin) angular momentum of the radiation field.  That
is, the hypothesis of locality is applied by the noninertial observer to the pointwise measurement of
the electromagnetic field; then, the result is subjected to the nonlocal process of Fourier analysis in
the stationary reference system of the uniformly rotating observer [28, 29]. 

To measure the frequency of a helicity component in the eikonal approximation, the noninertial observer
needs the record of field measurements for at least a few periods of the wave, i.e. $t\sim 2\pi
n/\omega$ for $n\geq 1$.  For high frequency waves with $\omega >>\Omega$, the effective rotation
frequency of the field about the direction of propagation of a wave of definite helicity is in this case
essentially $\omega$ as measured in the rotating frame over a timescale $t$ such that $\Omega t<< 1$. 
This result follows from expanding the cosine and sine functions in the expressions for ${\bf
E}\cdot{\bf\hat x}^{\prime}$ and ${\bf E}\cdot{\bf\hat y}^{\prime}$, respectively, in powers of $\Omega
t$ and keeping the lowest-order terms.  We note that $\omega^{\prime}=\omega$ is consistent with the
fact that ${\bf k} \cdot \mbox{\boldmath$\Omega$} = 0$ in this case and hence the helicity-rotation
coupling vanishes.  

For an arbitrary angle of incidence $\theta$, one may decompose $\mbox{\boldmath$\Omega$}$ into a
component of magnitude $\Omega\cos\theta$ parallel to the wave vector ${\bf k}$ and a component of
magnitude $\Omega\sin\theta$ perpendicular to {\bf k}.  Based on what has been discussed, we only expect
a frequency change due to the parallel component, i.e. $\omega^{\prime} = \omega
\mp\Omega\cos\theta$.  To make this heuristic argument more precise, we note that equation (5) is still
valid in the rotating frame, where the triad $({\bf\hat x}, {\bf\hat n}, {\bf\hat k})$ now refers to the
rotating axes, i.e.

\begin{eqnarray}  {\bf\hat x} &=& \cos \Omega t\; {\bf\hat x}^{\prime} - \sin\Omega t\;{\bf\hat
y}^{\prime}\;\;,\\  {\bf\hat n} &=& \cos\theta\;(\sin\Omega t\; {\bf\hat x}^{\prime} + \cos \Omega
t\;{\bf\hat y}^{\prime}) +
\sin\theta\;{\bf\hat z}^{\prime}\;\;,\\  {\bf\hat k} &=& -\sin\theta\;(\sin\Omega t\; {\bf\hat
x}^{\prime} +
\cos\Omega t\;{\bf\hat y}^{\prime}) +
\cos\theta\;{\bf\hat z}^{\prime}\;\;.
\end{eqnarray}

\noindent Since the observer axes rotate with frequency $\Omega$, vectors that are fixed in the inertial
frame precess in the opposite sense with respect to the noninertial observer.  It is therefore necessary
to define a new polarization basis $(\mbox{\boldmath$\xi$}, \mbox{\boldmath$\nu$})$ that remains
``fixed'' in the rotating frame as much as possible.  In this way, much of the temporal evolution of the
original basis with respect to the observer axes $({\bf\hat x}^{\prime}, {\bf\hat y}^{\prime}, {\bf\hat
z}^{\prime})$ is transferred to the phase of wave.  That is, the orthonormal triad
$({\mbox{\boldmath$\xi$}},{\mbox{\boldmath$\nu$}},{\bf\hat k})$ is related to $({\bf\hat x}, {\bf\hat
n}, {\bf\hat k})$ by a rotation of angle $\phi$ about the direction of wave propagation,

\begin{eqnarray} {\mbox{\boldmath$\xi$}} &=& \cos\phi\;{\bf\hat x} + \sin \phi\;{\bf\hat n}\;\;,\\
{\mbox{\boldmath$\nu$}} &=& -\sin\phi\;{\bf\hat x} + \cos\phi\;{\bf\hat n}\;\;,
\end{eqnarray}

\noindent such that ${\bf\hat x} \pm i{\bf\hat n} = (\mbox{\boldmath$\xi$}\pm i
\mbox{\boldmath$\nu$})\;{\rm exp}(\pm i \phi)$ and $\phi = 0$ at $t=0$ .  Combining equations (9) - (11)
and (12) - (13), we find by inspection that 

\begin{equation}
\sin \phi = \frac{1}{D}\cos\theta\sin\Omega t\;\;,\;\;\cos \phi = \frac{1}{D}\cos\Omega t\;\;,
\end{equation}

\noindent where $D>0$ is given by 

\begin{equation} D^2 = \cos^2\theta + \sin^2\theta\cos^2\Omega t\;\;.
\end{equation}

\noindent Let us note that relations (14) for $\phi$ reduce to $\phi = \Omega t$ for $\theta = 0, \phi =
-\Omega t$ for $\theta = \pi$ and $\phi =0$ for $\theta = \pi/2$, as expected.  According to the
noninertial observer, the electric field

\begin{equation} {\bf E}(t, {\bf 0}) = {\rm Re}[A(\mbox{\boldmath$\xi$} \pm
i\mbox{\boldmath$\nu$})e^{-i\omega t\pm i\phi}]
\end{equation}

\noindent represents a circularly polarized wave with a new phase $\Phi =-\omega t\pm\phi$.  The
frequency of the wave is given in the eikonal approximation by $-\partial\Phi/\partial t = \omega
\mp\partial\phi/\partial t$, where $\partial\phi/\partial t = D^{-2}\Omega\cos\theta$ by equations (14)
- (15).  The measurement of the frequency would require observation of at least a few oscillations of
the wave.  Over such a length of time $t$, starting from $t=0$, we have
$\epsilon = \Omega t << 1$ since $\omega >>\Omega$.  It follows from equation (15) that $D^{-2} = 1 +
\epsilon^{2}\sin^2\theta + O(\epsilon^4)$.  Therefore, $\omega^{\prime} = \omega \mp\Omega\cos\theta =
\omega - {\bf\hat H}\cdot\mbox{\boldmath$\Omega$}$ in the eikonal approximation under consideration here.

Our discussion has been based on beams of radiation that are initially circularly polarized.  However,
we could have started with an incident linearly polarized beam that may be considered a coherent
superposition of positive and negative helicity waves.  Our result that for these states
$\omega^{\prime} = \omega\mp\Omega\cos\theta$ is equivalent to the intuitive expectation that from the
viewpoint of the rotating observer, the direction of linear polarization---fixed in the inertial
frame---must precess in the opposite sense [13].

More generally, for an electromagnetic radiation field received by an observer rotating uniformly with
frequency $\Omega$ about the $z$-axis on a circle of radius $r$, the Fourier analysis of the measured
field implies that $\omega^{\prime} = \gamma (\omega - M\Omega)$, where $M = 0, \pm 1, \pm 2, ...$ is
the multipole parameter such that $\hbar M$ is the total angular momentum of the field along the
$z$-axis.  In the eikonal approximation $\omega >> \Omega, \hbar\omega^{\prime}\approx\gamma(\hbar\omega
- {\bf J} \cdot\mbox{\boldmath$\Omega$})$, where the total angular momentum ${\bf J}$ is a sum of
orbital plus spin contributions, i.e. ${\bf J} = {\bf L} + {\bf S}$.  Thus with ${\bf L} = {\bf
r}\times{\bf p}, {\bf p} = \hbar{\bf k}$ and ${\bf v} = \mbox{\boldmath$\Omega$}\times{\bf r}$, we find
that $\omega^{\prime}\approx\gamma(\omega - {\bf v}\cdot{\bf k}) - \gamma
{\bf\hat H}\cdot\mbox{\boldmath$\Omega$}$; in this way, we recover the Doppler effect together with the
spin-rotation coupling [28, 29].  Neglecting time dilation, the relation $\omega^{\prime} = \omega -
M\Omega$ has recently received experimental confirmation in the work of Courtial et al. [30, 31], who
demonstrated that $M$ is the {\it total} angular momentum parameter [31].  Further interesting
discussions of this issue are contained in [32, 33].
\vspace{.25in}

\noindent 3. Doppler tracking of spacecraft
\vspace{.05in}

Circularly polarized radio beams are routinely employed in Doppler tracking of spacecraft [7, 8].  The
helicity-rotation coupling must be taken into account whenever rotating emitters and/or receivers are
involved.  As an example, let us consider an emitter of velocity ${\bf v}_1$ and rotation frequency
$\mbox{\boldmath$\Omega$}_1$ sending a signal of definite helicity to a receiver of velocity ${\bf v}_2$
and rotation frequency $\mbox{\boldmath$\Omega$}_2$.  We neglect plasma effects and assume that gravity
is turned off.  The emitter and receiver are then isolated systems that move in a global inertial
frame.  Let $P$ be a point along the beam and consider an inertial observer at rest at $P$ in the
inertial frame.  Then the modified Doppler formula connecting the frequency $\omega_1$, measured by a
noninertial observer comoving with the emitter, with $\omega$ measured at $P$ is 

\begin{equation}
\omega_1 = \gamma_1(1-\mbox{\boldmath$\beta$}_1\cdot{\bf\hat k})\omega\mp\gamma_1{\bf\hat
k}\cdot\mbox{\boldmath$\Omega$}_1\;\;.
\end{equation}

\noindent Similarly, the modified Doppler formula connecting $\omega$ with the frequency of the wave
$\omega_2$ measured by a noninertial observer comoving with the receiver is 

\begin{equation}
\omega_2 = \gamma_2(1-\mbox{\boldmath$\beta$}_2\cdot{\bf\hat k})\omega\mp\gamma_2{\bf\hat
k}\cdot\mbox{\boldmath$\Omega$}_2\;\;.
\end{equation}

\noindent Combining equations (17) and (18) so as to eliminate $\omega$, we find

\begin{equation}
\frac{\sqrt{1-\beta_1^2}\;\omega_1 \pm{\bf\hat
k}\cdot\mbox{\boldmath$\Omega$}_1}{\sqrt{1-\beta_2^2}\;\omega_2\pm{\bf\hat
k}\cdot\mbox{\boldmath$\Omega$}_2}\;=\;\frac{1-\mbox{\boldmath$\beta$}_1\cdot{\bf\hat
k}}{1-\mbox{\boldmath$\beta$}_{2}\cdot{\bf\hat k}}\;\;.
\end{equation}

\noindent In equations (17) - (19), $\mbox{\boldmath$\beta$} = {\bf v}/c$ and the upper (lower) sign
refers to positive (negative) helicity radiation; moreover, all plasma and gravity effects are neglected.

The influence of gravitation on the propagation of electromagnetic rays has been extensively studied
(see [34] and the reference cited therein); therefore, we limit our considerations here to wave effects
associated with the helicity of the radiation.  One can show that helicity is in general conserved for
radiation propagating in vacuum in a gravitational field; moreover, positive helicity radiation is
scattered differently than negative helicity radiation due to the coupling of the helicity with the
gravitomagnetic field [35, 36].  This coupling, which can lead to differential deflection of polarized
radiation, is the gravitational analog of the helicity-rotation coupling and can be estimated via the
gravitational Larmor theorem [37].  For instance, for an emitter fixed on a rotating gravitational
source (such as the Earth) with angular momentum ${\bf J}$ and rotation frequency
$\mbox{\boldmath$\Omega$}$, the effective frequency of rotation in the helicity-rotation coupling becomes
$\mbox{\boldmath$\Omega$} - \mbox{\boldmath$\Omega$}_{\rm P}$ in the linear approximation of general
relativity, where 

\begin{equation}
\mbox{\boldmath$\Omega$}_{\rm P} = \frac{GJ}{c^2r^3}[3({\bf\hat J}\cdot{\bf\hat r}){\bf\hat r} - {\bf\hat
J}]
\end{equation}

\noindent is the precession frequency of a local gyroscope in the gravitomagnetic field of the source. 
In all cases of interest within the solar system, the helicity-gravitomagnetic field coupling is at most
about a million times smaller than the helicity-rotation coupling.
\vspace{.25in}

\noindent 5. Discussion
\vspace{.05in}

For an observer rotating uniformly with frequency $\mbox{\boldmath$\Omega$}$, an incident
electromagnetic wave of frequency $\omega$ is expected to have a spectrum of frequencies
$\omega^{\prime} = \gamma(\omega - M\Omega)$ with $M=0, \pm 1, \pm 2, ...$.  In the eikonal
approximation, $\omega^{\prime}$ may be written in the form of equation (1).  For radiation propagating
along the direction of rotation of the observer, the exact result is $\omega^{\prime} =
\gamma(\omega\mp\Omega)$ depending on the helicity of the incident radiation.  Thus the rotation of the
observer affects the phase of the wave, but not its amplitude.  This is consistent with all of the
observational data discussed thus far; that is, the helicity amplitudes are independent of the rotation
of the observer.  

These results follow from the analysis of the electromagnetic field as measured pointwise by the
rotating observer.  This standard theory, based on the hypothesis of locality, thus contains the
theoretical conclusion that---under certain circumstances that have not been accessible
experimentally thus far---$\omega^{\prime} = 0$, so that the electromagnetic field would stand completely
still as measured by the rotating observers.  For instance, this would be the case if the observers
rotate with frequency $\Omega = \omega$ in the positive sense about the direction of propagation of a
plane RCP wave of frequency $\omega$.  Thus by a mere rotation an electromagnetic wave could be made
to stand completely still with respect to the rotating observers; in fact, there is no observational
evidence at present in favor of the occurrence of such a phenomenon.  This situation is analogous to the
status of the pre-relativistic Doppler formula : For an observer moving with $v << c$, it was in
reasonable agreement with observation, but for $v=c$, which was not accessible to observation,
$\omega^{\prime}$ could vanish.  The theory of relativity avoids this conclusion, since a particle with
nonzero mass is forbidden from ever reaching the speed of light.

The theory of relativity thus expressly prohibits $\omega^{\prime}=0$ for inertial observers in
Minkowski spacetime and in the eikonal {\it limit} of ray propagation for arbitrary accelerated
observers in a gravitational field.  To generalize this aspect of the standard theory, a nonlocal theory
of accelerated systems has been developed [38-40] that goes beyond the hypothesis of locality and is
consistent with all observational data available to date.  The nonlocal theory takes the past history
of a noninertial observer into account and is based on the {\it assumption} that a radiation field
can never stand completely still with respect to any observer; therefore, it is very important to
subject this theory to direct experimental tests.   In this theory, $\omega^{\prime} =
\gamma(\omega - M\Omega)$ as before except for the case of ``resonance'' at $\omega = M\Omega$, where the
electromagnetic field is no longer constant in time but exhibits resonance-type behavior.  Moreover, as
a direct consequence of nonlocality, the helicity-rotation coupling is also reflected in the helicity
amplitudes.  For instance, for circularly polarized radiation of frequency $\omega >>\Omega$ propagating
along the axis of rotation of the observer, the measured amplitude of the wave corresponding to the
fields rotating in the same (opposite) sense as the observer is enhanced (diminished) by a factor of
$1+\Omega/\omega\:(1-\Omega/\omega)$.  Thus the predicted relative change in the intensity response is
$2\Omega/\omega\approx 2 \times 10^{-8}$ for the GPS phase wrap-up experiment reported in [8], while for
the microwave experiment of Allen [10], $2\Omega/\omega\approx 4 \times 10^{-9}$.  In the optical
regime, $2\Omega/\omega$ would be even much smaller, as expected.  It would be very interesting to test
this prediction of the nonlocal theory using a fast spinning radio receiver by searching for the
coupling of helicity with rotation in the intensity response of the receiver.

\vspace{.15in}
\noindent Acknowledgments

I am grateful to John D. Anderson and Neil Ashby for helpful correspondence.  Thanks are also due to
Sergei Kopeikin for interesting discussions.

\newpage

\noindent References\\
\baselineskip18pt
\vspace{.05in}
\begin{description}
\item{[1]} A. Einstein, Ann. Phys. (Leipzig) 17 (1905) 891.
\item{[2]} B. Mashhoon, Phys. Lett. A 122 (1987) 67.
\item{[3]} B. Mashhoon, Phys. Lett. A 139 (1989) 103.
\item{[4]} B. Mashhoon, Gen. Rel. Grav. 31 (1999) 681.
\item{[5]} B. Mashhoon, Class. Quantum Grav. 17 (2000) 2399.
\item{[6]} J.D. Anderson et al., Phys. Rev. Lett. 81 (1998) 2858.
\item{[7]} J.D. Anderson et al., Phys. Rev. D 65 (2002) 082004.
\item{[8]} N. Ashby, in: N. Dadhich and J. Narlikar (eds.), Gravitation and Relativity: At the Turn of
the Millennium, Proceedings of GR-15 (Inter-University Center for Astronomy and Astrophysics, Pune,
India, 1997), pp. 231-258.
\item{[9]} N. Ashby, Physics Today 55 (May, 2002) 41.
\item{[10]} P.J. Allen, Am. J. Phys. 34 (1966) 1185.
\item{[11]} R.A. Beth, Phys. Rev. 48 (1935) 471.
\item{[12]} R.A. Beth, Phys. Rev. 50 (1936) 115.
\item{[13]} B. Mashhoon, R. Neutze, M. Hannam and G.E. Stedman, Phys. Lett. A 249 (1998) 161.
\item{[14]} R.G. Newburgh and G.V. Borgiotti, Appl. Opt. 14 (1975) 2727.
\item{[15]} R. Crane, Appl. Opt. 8 (1969) 538.
\item{[16]} G.E. Sommargren, J. Opt. Soc. Am. 65 (1975) 960.
\item{[17]} R.N. Shagam and J.C. Wyant, Appl. Opt. 17 (1978) 3034.
\item{[18]} H.Z. Hu, Appl. Opt. 22 (1983) 2052.
\item{[19]} M.P. Kothiyal and C. Delisle, Opt. Lett. 9 (1984) 319.
\item{[20]} J.P. Huignard and J. P. Herriau, Appl. Opt. 24 (1985) 4285.
\item{[21]} B.A. Garetz and S. Arnold, Opt. Commun. 31 (1979) 1.
\item{[22]} B.A. Garetz, J. Opt. Soc. Am. 71 (1981) 609.
\item{[23]} R. Simon, H.J. Kimble and E.C.G. Sudarshan, Phys. Rev. Lett. 61 (1988) 19.
\item{[24]} F. Bretenaker and A. Le Floch, Phys. Rev. Lett. 65 (1990) 2316.
\item{[25]} R. Bhandari, Phys. Rep. 281 (1997) 1.
\item{[26]} V. Bagini et al., Eur. J. Phys. 15 (1994) 71.
\item{[27]} A.B. Pippard, Eur. J. Phys. 15 (1994) 79.
\item{[28]} B. Mashhoon, Found. Phys. 16 (Wheeler Festschrift) (1986) 619.
\item{[29]} L.H. Ryder and B. Mashhoon, gr-qc/0102101.
\item{[30]} J. Courtial et al., Phys. Rev. Lett. 80 (1998) 3217.
\item{[31]} J. Courtiel et al., Phys. Rev. Lett. 81 (1998) 4828.
\item{[32]} G. Nienhuis, Opt. Commun. 132 (1996) 8.
\item{[33]} I. Bialynicki-Birula and Z. Bialynicka-Birula, Phys. Rev. Lett. 78 (1997) 2539.
\item{[34]} S. Kopeikin and B. Mashhoon, Phys. Rev. D 65 (2002) 064025.
\item{[35]} B. Mashhoon, Nature 250 (1974) 316.
\item{[36]} B. Mashhoon, Phys. Rev. D 11 (1975) 2679.
\item{[37]} B. Mashhoon, Phys. Lett. A 173 (1993) 347.
\item{[38]} B. Mashhoon, Phys. Rev. A 47 (1993) 4498.
\item{[39]} C. Chicone and B. Mashhoon, Ann. Phys. (Leipzig) 11 (2002) 309.
\item{[40]} C. Chicone and B. Mashhoon, Phys. Lett. A 298 (2002) 229.
\end{description}

\end{document}